\documentclass[conference]{IEEEtran}
\IEEEoverridecommandlockouts
\usepackage{biblatex}
\usepackage{amsmath,amssymb,amsfonts}
\usepackage{algorithmic}
\usepackage{graphicx}
\usepackage{textcomp}
\usepackage{xcolor}
\usepackage{enumitem}
\usepackage{multirow}
\usepackage{adjustbox}
\usepackage[hidelinks]{hyperref}

\usepackage{nimbusmononarrow} % Use a more narrow monospace font
\usepackage[T1]{fontenc}

\def\BibTeX{{\rm B\kern-.05em{\sc i\kern-.025em b}\kern-.08em
    T\kern-.1667em\lower.7ex\hbox{E}\kern-.125emX}}
\begin{document}

% \title{Designing Recommendation Systems for Interface Improvement

\title{Towards the Assisted Decomposition of Large-Active Files
}

\author{\IEEEauthorblockN{Jason Lefever}
\IEEEauthorblockA{\textit{Drexel University}\\
Philadelphia, USA \\
jtl86@drexel.edu}
\and
\IEEEauthorblockN{Yuanfang Cai}
\IEEEauthorblockA{\textit{Drexel University}\\
Philadelphia, USA \\
yc349@drexel.edu}
\and
\IEEEauthorblockN{Rick Kazman}
\IEEEauthorblockA{\textit{University of Hawaii}\\
Honolulu, USA \\
kazman@hawaii.edu}
\and
\IEEEauthorblockN{Hongzhou Fang}
\IEEEauthorblockA{\textit{Drexel University}\\
Philadelphia, USA \\
hf92@drexel.edu}
% \and
% \and
% \and
}

\maketitle

% Also, papers must include in the abstract and the introduction a clear statement about the claimed contribution, i.e., “New Ideas”or “Emerging Results”.
\begin{abstract}
Tightly coupled and interdependent systems inhibit productivity by requiring developers to carefully coordinate their changes, even when modifying subsystems that \emph{should} be independent of one another. Poor architectural decisions frequently lead to the presence of large, change-prone source files that are at the center of complexes of technical debt~\cite{Mo-2021-ArchitectureAnti-Pa, Lefever-2021-OntheLackofConse}. This kind of technical debt quickly incurs interest both through coordination costs and error-proneness.
In this paper, we present a new approach for decomposing these ``large-active'' files to pay down critical technical debt. We package our approach as a refactoring recommendation system.  Each recommendation is determined by analyzing patterns of co-change and mutual dependency among files. Each recommendation corresponds to a responsibility the large-active file has in relation to the rest of the system. By moving recommended functionality from the large-active file into smaller files, developers can reduce the impact of a debt-laden file and clarify its essential responsibilities. A key advantage of this approach over prior work is that we better focus effort; we avoid spending developer effort refactoring code that is only superficially problematic. We achieve this by incorporating revision history into both determining and ranking recommendations. Each recommendation corresponds to some change-prone responsibility. We present some  examples of this approach in action and outline our future plans.

\end{abstract}


\section{Introduction}\label{introduction}
% parallel and independent changes 
% novelty of using co-change for refactoring
% unstable interface and crossing are the means of improving parallelism

% - large, active, error-prone
% - influence
% - influence comes from structure and history
% Oversized and change-prone source files are a significant contributor to a system's error-proneness~\cite{Mo-2021-ArchitectureAnti-Pa} and are the most reliable predictor of future technical debt~\cite{Lefever-2021-OntheLackofConse}. The \emph{unstable interface} and \emph{crossing} anti-patterns have been introduced to explain this correlation as the effect of diverging responsibilities.

% In this paper, we present our method of algorithmically decomposing large files and offering these decompositions to the developers as refactoring recommendations.

% cleanly dividing the responsabilities of a system into a dividing tasks among developers.
% In his canonical essay \emph{The Mythical Man Man-Month}, Fred Brooks scrutinizes the common assumption that assigning more developers to a project will necessarily speed up its development. The core problem that teams face is that most tasks are not easily divisible among developers.
The difficulty of modularizing a system so that developers can work independently of one another is as old as software development~\cite{Parnas-1972-Onthecriteriatob}. While some development tasks are inherently sequential, even those that \textit{should} be parallel often turn out to be coupled in frustrating ways. A common place where developers discover this coupling is deep within the bowels of large, ever-growing, and frequently changing source files. Because of their extensive influence on the rest of the system, these large-active files often need to be changed for seemingly unrelated reasons. Developers working on otherwise parallel tasks must therefore carefully coordinate with one another to successfully modify such files. These files are often identified by smells such as god file~\cite{fowler:refactor99}, unstable interface~\cite{Mo-2021-ArchitectureAnti-Pa}, or crossing~\cite{Mo-2021-ArchitectureAnti-Pa}. Recently, this observation was given empirical backing when the presence of large-active files was shown to the strongest indicator of technical debt~\cite{Lefever-2021-OntheLackofConse}. In this paper, we present our novel method of algorithmically recommending operations that can modularize a large-active file to increase the overall parallelism of the development team.

% decomposing large-active files. These decompositions are  and offering these decompositions as refactoring recommendations.

% These large-active files have shown to be the strongest indicator of technical debt and 

% Excessively large and change-prone files

% Thus slowing development by inhibiting the parallelism of development.

% {\color{orange} A \emph{large file} (or when appropriate, \emph{large class}) is a source file within a software system that is ``trying to do too much''~\cite{fowler:refactor99,Fowler-2018-Refactoring}. Not only do large files violate widely held best practices like the single responsibility principle (SRP)~\cite{Martin-2000}, they have also been shown to be a significant contributor to a system's error-proneness~\cite{Mo-2021-ArchitectureAnti-Pa} and the most reliable predictor of future technical debt~\cite{Lefever-2021-OntheLackofConse}. The related anti-patterns of \emph{unstable interface} and \emph{crossing} have been introduced to explain this error correlation as the effect of diverging responsibilities. Even when working on seemingly unrelated modules, these large files must frequently be changed as they are bound up with so many different concerns.} In this paper, we present our method of algorithmically decomposing large files and offering these decompositions to the developers as refactoring recommendations.

% While some may have started small, these files now occupy a significant number of responsibilities within their system.

No existing refactoring recommendation system is able to directly eliminate a large-active file. Instead, most existing refactoring recommendation systems function by recommending operations that improve quality metrics~\cite{Harman-2007-Paretooptimalsearc,Ouni-2013-Maintainabilitydefe,Mkaouer-2014-Recommendationsyste,Mkaouer-2015-Many-ObjectiveSoftw,Alizadeh-2019-RefBotIntelligent}. These systems use quality metrics to measure global characteristics like coupling, cohesion, maintainability, etc. These measurements are then used as the targets of a multi-objective search algorithm to find a sequence of refactoring steps. However, no particular attention is paid to large-active files despite 
these often being the focal point of maintenance activities. In fact, no attention is paid at all to the \emph{activity} or \emph{co-activity} of files as change history is ignored by these existing systems. So while these systems will improve their selected quality metrics, and while some of these quality metrics will encourage smaller files, it is not likely that these systems will \emph{reliably} decompose a large file into its constitutive parts, let alone a large-active file. Furthermore, because change history is not considered, it is possible that  these approaches are not prioritizing code that is incurring the most effort and the greatest interest. Given the recent evidence suggesting that these files play a central role in the worsening of technical debt, it is especially important to have  algorithms aimed specifically at the decomposition of excessively large and active files.

% to the local structure surrounding such files.

%  Crucially, none of these systems considers the activity or co-activity of files

%  change history, crucial information about the activity of files  is being ignored

% it is doubtful that they can reliably decompose large files given that they were not built for this task. This is especially important considering recent results that suggest large files play a central role in worsening technical debt.

% select quality metrics will be improved, it is improbable that any large files have been successfully decomposed. T

% While the goal of decomposing a large file and the goal of improving certain quality metrics is not entirely orthogonal

A large-active file with many dependents is problematic because of the extensive influence it has on the rest of the system that prevents parallel changes and inhibits modularity. This influence can be seen in the number of  files dependent on it (direct or otherwise), and in the number of times it has co-changed with other files. Our approach is a form of ensemble clustering that uses these shared influences to calculate the similarity between members of the target file. The resulting clusters are served as recommendations to reduce the influence that this file has on the rest of the system. These refactoring recommendations identify clear responsibilities of the large-active file and guide a developer to pull these out into their own file. When a developer refactors according to her choice of recommendation, she is reducing the influence of the target file while also narrowing its actual responsibilities to only the most essential.

% Our approach examines the extensive influence a large-active file has on the rest of the system to determine promising locations where it could be broken apart.

% Our approach exploits the structure of an unstable interface or crossing to provide recommendations that can directly decompose a large file. These anti-patterns inform us that a large file is problematic because of its outsized level of influence. This influence includes both other files that are tightly coupled with the large file and the periods of revision history in which they have changed together. Our proposed algorithm identifies subsets of the target file that share common influences and offers them as a ranked list to the developer. Each recommended subset represents an unambiguous responsibility of the target file as suggested by its influences. So when a developer refactors according to her choice of these recommendations, she is both reducing the influence of the target file while narrowing its responsibilities to the most essential.

The promise of our approach is already evident in our early results. These results indicate that our algorithm successfully recommends cohesive clusters of functionality, each associated with a distinct responsibility of the target file. Specifically, we have found the most promise in ``utility'' files that span many responsibilities of the system. These kinds of files are easy to evaluate because of their deliberate nature. Still, our goal with this approach is to find more subtle distinctions that a developer might miss. This evaluation is left as future work.

This work is significant because it is the first to offer a way to directly decompose large-active source files, improving their cohesion, thus enabling parallel developer activity and reducing their unwarranted (bug-producing, churn producing) complexity. Large-active source files have repeatedly been shown to correlate with technical debt and software decay. This correlation stems for the extensive influence they have on the rest of the system. Our approach promises to reverse this influence, decreasing the influence of a large-active file on the rest of the system.

\section{Approach}\label{approach}
Our refactoring recommendation approach is designed to disentangle large-active files. An excellent formulation of the problem is the unstable interface anti-pattern of Mo et al~\cite{Mo-2021-ArchitectureAnti-Pa}. An \emph{unstable interface} is a file with many dependents that also changes frequently with those dependents. We identify two core problems any unstable interface will exhibit.
\begin{enumerate}[label=\alph*)]
\item \emph{A poor separation of concerns.} This file must be changed for many different reasons. Other files use this file for many different reasons.
\item \emph{A poor separation between interface and client.} This file must be changed with many of its clients in the same commits. This implies that a poor abstraction boundary was drawn between components. Implementation details of the unstable interface may be leaking into the clients.
\end{enumerate}
We address these two problems separately using distinct algorithms. Then we aggregate and rank the recommendations from both algorithms. The top recommendations are given to the developer.

% We identify two core problems any unstable interface must exhibit: (a) a poor separation of concerns -- this file has too many responsibilities, and (b) a leaky abstraction -- this file is leaking details of its implementation to clients. We address these two problems separately using distinct algorithms. Then we aggregate and rank the recommendations from both algorithms. The top recommendations are given to the developer.

% There are two primary problems 

% A \emph{crossing} is an unstable interface that additionally has many dependees and changes frequently with those as well. We introduce two algorithms 

% Our approach consists of two general algorithms: \emph{interface splitting} and \emph{interface redrawing}. Both algorithms produce a number of refactoring recommendations which are then aggregated together and ranked. The top recommendations are given to the developer.

\subsection{Interface Splitting}
The goal of the interface splitting algorithm is to reduce the responsibilities of the target file to its most essential by pulling other responsibilities out and placing them in their own file.

% The target file is represented as a set of its functions. Each recommendation is a subset of its functions that should be pulled out into its own stand-alone file.

% The \emph{interface splitting} algorithm identifies subsets of the target file that serve specific responsibilities by analyzing how it influences the rest of the system. These subsets are recommended to the developer to be extracted into their own stand-alone file.

Let $T$ be the set of functions found in the target file and $D$ be the set of files that are dependent on the target file. Then for any target function $t \in T$, let $d(t) \subseteq D$ denote the subset of dependents that use that function. This dependency could be a direct function call, a transitive function call, or a co-change dependency mined from the revision history.

To determine if two functions, $t_i$ and $t_j$, serve the same responsibilities, we check the similarity of their respective dependent sets, $d(t_i)$ and $d(t_j)$. The dependent set of any particular function captures the responsibilities it has to the rest of the system. Let the Jaccard index ${s_{ij}=\lvert d(t_i) \cap d(t_j) \rvert/\lvert d(t_i) \cup d(t_j) \rvert}$
% $$
% % s_{ij}=\lvert c(f_i) \cap c(f_j) \rvert/\lvert c(f_i) \cup c(f_j) \rvert
% s_{ij}=\frac{\lvert c(f_i) \cap c(f_j) \rvert}{\lvert c(f_i) \cup c(f_j) \rvert}
% $$
be this similarity. Notice that the more clients these two functions have in common, the higher~$s_{ij}$ is, and the fewer they have in common, the lower~$s_{ij}$ is. So pairs of functions with a high~$s_{ij}$ are considered to play a similar role in the wider system.

% So target functions with similar responsibilities

% This fraction measures the similarity of the responsibilities these two functions have to the rest of the system.

% influences (function calls, co-changes, etc.)

Then, to discover subsets of $T$ that may serve as cohesive stand-alone files, we cluster the rows (or equivalently, columns) of the square similarity matrix $\mathbf{S}_s=[s_{ij}]$ and take each resulting cluster as a candidate recommendation. Functions that serve similar sets of clients should be grouped together, while functions that serve not-so-similar sets of clients should be separate. There are many clustering algorithms of this sort in the literature~\cite{Xu-2015-AComprehensiveSurv}, but for now, we simply assume that whichever algorithm is chosen is parameterized with some scalar or vector value $\mathbf{\theta}$. For instance, $\mathbf{\theta}$ may be the desired number of clusters.

% if using $k$-means, or maybe \texttt{MinPts} and \texttt{Eps} if using DBSCAN.

Because we do not want the developer to have to specify the parameter themselves, we require the use of a heuristic function $g_q(\mathbf{S}_s)=\theta_q$ that ``guesses'' a good parameter for clustering the similarity matrix, where $q=1$ is the highest quality guess, $q=2$ is the second highest, etc. Examples of~$g$ include the silhouette coefficient~\cite{ROUSSEEUW198753} and the spectral gap~\cite{vonLuxburg2007}.

The interface splitting algorithm proceeds as an ensemble clustering algorithm~\cite{pons:survey-ensemble}. First, the similarity matrix $\mathbf{S}_s$ is calculated. Then, the best guess parameter $\theta_1$ is derived from the similarity matrix. The clustering algorithm is executed with $\theta_1$ and the resulting clusters are added to the multiset of recommendations $R$. Then, the next-best guess $\theta_2$ is calculated, $\mathbf{S}_s$ is re-clustered, and the clusters are added to $R$. This process continues until $q=q_{\text{max}}$. Finally, each recommendation of $R$ is ranked, first by multiplicity then by average change frequency, both in descending order.

% is presented to the developer in descending order of average change-frequency.

% Then the clustering algorithm is executed again with $\theta_2$ and the results are 

% but for this discussion we simply assume that 

% There are many clustering algorithms of this sort in the literature, but for our experiments we used a variant of spectral clustering based on normalized cut, given its apparent success across many domains.

% Notice that $S_{ij}=1$ when these functions serve identical sets of clients, $S_{ij}=0$ when these functions serve disjoint sets of clients, and $S_{ij}$ is somewhere in-between otherwise.

% and $z : F \to 2^D$ be a map a function to the set of files that call it. So if $z(f_3)=\{d_1, d_4, d_5\}$ then $f_3$ is called exclusively by $d_1$, $d_4$, and $d_5$. We also include transitive calls in $z$. So $d_1$ may not directly call $f_3$, but rather $d_1$ may call $f_1$ which calls $f_2$ which calls $f_3$. In this case, $d_1 \in z(f_3)$ as long as all intermediate functions are also inside $F$.

% Let $S$ be the similarity matrix where each cell $S_{f_if_j}$ encodes how similar the function $f_i$ is with $f_j$ and vice versa. A good choice for $S_{f_if_j}$ is the Jaccard index~\cite{jaccard,costa:jaccard-index} between $z(f_i)$ and $z(f_j)$. Let
% $$
% S_{f_if_j} = \frac{\lvert z(f_i) \cap z(f_j) \rvert}{\lvert z(f_i) \cup z(f_j) \rvert}.
% $$
% So $S_{f_if_j}$ will be $1$ if $f_i$ and $f_j$ are called by the exact same dependent files, 0 if they share no dependent files, or somewhere in between if they share some dependent files.

\subsection{Interface Redrawing}
The goal of the interface redrawing algorithm is to alter the boundary between the target file and its clients so that future changes are less likely to modify both sides of the interface. The interface redrawing algorithm is similar to the interface splitting algorithm but instead of only considering functions of the target file, functions of the client file are also considered. The algorithm recommends subsets containing functions from both sides of the interface to the developer. A recommended subset may then either be extracted into its own file so it is a new client of the interface, or it may be moved into the target file so it is ``behind'' the interface.

Let $C$ be the set of commits found in this project and~${c(f) \subseteq C}$ be the set of commits where the function~${f \in (T \cup D)}$ changed. Let $\mathbf{S}_r = [s_{ij}]$ be a rectangular similarity matrix where rows correspond to target functions and columns correspond to client functions. Each entry is the Jaccard index ${s_{ij}=\lvert c(t_i) \cap c(d_j) \rvert/\lvert c(t_i) \cup c(d_j) \rvert}$. This is the similarity between the change history of the target function~$t_i$ and the client function~$d_j$. It does not matter if~$d_j$ actually calls~$t_i$ because their co-change relationship alone indicates that they depend on one another. We conjecture that these co-change relationships reveal locations where implementation details of the target file are being leaked to clients.

By clustering the rows and columns of $\mathbf{S}_r$ at the same time we discover portions of the interface that are prone to instability and leakage. Each resulting cluster contains functions from both sides of the interface that tend to change together. In general, this form of clustering is known as co-clustering~\cite{dhillon2001co}. As before, we sidestep discussion of specific co-clustering algorithms and instead focus on the parameter $\theta$. Here, we also require a heuristic function ${g_q(\mathbf{S}_r)=\theta_q}$ that can provide promising parameters for co-clustering the similarity matrix.

The interface redrawing algorithm proceeds as an ensemble clustering algorithm~\cite{pons:survey-ensemble}. First, the similarity matrix $\mathbf{S}_r$ is calculated. Then, the best guess parameter $\theta_1$ is derived from the similarity matrix. The co-clustering algorithm is executed with $\theta_1$ and the resulting co-clusters are added to the multiset of recommendations $R$. Then, the next-best guess $\theta_2$ is calculated, $\mathbf{S}_r$ is re-clustered, and the co-clusters are added to $R$. This process continues until $q=q_{\text{max}}$. Finally, each recommendation of $R$ is ranked, first by multiplicity then by average change frequency, both in descending order.

\begin{table}[]
\caption{Interface Splitting -- Selected Recommendations}
\label{table:is-recs}
\centering
\begin{tabular}{|c|l|}
\hline
\multicolumn{1}{|l|}{\textbf{}} & \textbf{Recommendation}                               \\ \hline
\multirow{2}{*}{1}              & \texttt{Utils.java  > getApplicationLabel}            \\
                                & \texttt{Utils.java  >  getBadgedIcon}                 \\ \hline
\multirow{5}{*}{3}              & \texttt{Utils.java  >  getFaceManagerOrNull}          \\
                                & \texttt{Utils.java  >  getFingerprintManagerOrNull}   \\
                                & \texttt{Utils.java  >  hasFaceHardware}               \\
                                & \texttt{Utils.java  >  hasFingerprintHardware}        \\
                                & \texttt{Utils.java  >  isMultipleBiometricsSupported} \\ \hline
\multirow{2}{*}{6}              & \texttt{Utils.java  >  getCredentialOwnerUserId}      \\
                                & \texttt{Utils.java  >  getUserIdFromBundle}           \\ \hline
\multirow{2}{*}{9}              & \texttt{Utils.java  >  getSecureTargetUser}           \\
                                & \texttt{Utils.java  >  hasMultipleUsers}              \\ \hline
\multirow{2}{*}{10}             & \texttt{Utils.java  >  getBatteryPercentage}          \\
                                & \texttt{Utils.java  >  isBatteryPresent}              \\ \hline
\multirow{5}{*}{12}             & \texttt{Utils.java  >  createAccessibleSequence}      \\
                                & \texttt{Utils.java  >  enforceSameOwner}              \\
                                & \texttt{Utils.java  >  getCredentialOwnerUserId}      \\
                                & \texttt{Utils.java  >  getCredentialType}             \\
                                & \texttt{Utils.java  >  getUserIdFromBundle}           \\ \hline
\end{tabular}
\end{table}

\begin{table}[]
\caption{Interface Splitting -- Dependent Files of No. 3 in Table~I with \#~of Occurrences}
\label{table:is-clientset}
\centering
\begin{tabular}{|l|l|}
\hline
\textbf{File}                                                & \textbf{\#} \\ \hline
\texttt{CombinedBiometricStatusPreferenceController.java}    & 4           \\
\texttt{BiometricSettingsAppPreferenceController.java}       & 3           \\
\texttt{FaceSettings.java}                                   & 3           \\
\texttt{FaceStatusPreferenceController.java}                 & 3           \\
\texttt{FingerprintEnrollFinish.java}                        & 3           \\
\texttt{FingerprintEnrollSuggestionActivity.java}            & 3           \\
\texttt{FingerprintStatusPreferenceController.java}          & 3           \\
\texttt{BiometricFaceStatusPreferenceController.java}        & 2           \\
\texttt{BiometricFingerprintStatusPreferenceController.java} & 2           \\
\texttt{BiometricsSettingsBase.java}                         & 2           \\
\texttt{FaceEnrollIntroduction.java}                         & 2           \\
\texttt{FaceSettingsAppPreferenceController.java}            & 2           \\
\texttt{FaceSettingsKeyguardPreferenceController.java}       & 2           \\
\texttt{FingerprintEnrollIntroduction.java}                  & 2           \\
\texttt{FingerprintSuggestionActivity.java}                  & 2           \\
\texttt{ChooseLockGeneric.java}                              & 2           \\
\texttt{SetNewPasswordController.java}                       & 2           \\ \hline
\end{tabular}
\end{table}

\begin{table}[]
\caption{Interface Redrawing -- Selected Recommendations}
\label{table:ir-recs}
\centering
\begin{tabular}{|c|l|}
\hline
\multicolumn{1}{|l|}{\textbf{}} & \textbf{Recommendation}                                        \\ \hline
\multirow{2}{*}{1}              & \texttt{Utils.java  >  getCredentialType}                      \\
                                & \texttt{ConfirmDeviceCredentialActivity.java  >  onCreate}     \\ \hline
\multirow{2}{*}{6}              & \texttt{Utils.java  >  getSecureTargetUser}                    \\
                                & \texttt{SetNewPasswordController.java > launchChooseLock}      \\ \hline
\multirow{2}{*}{7}              & \texttt{Utils.java  >  isDemoUser}                             \\
                                & \texttt{FactoryResetPreferenceController.java  >  isAvailable} \\ \hline
\multirow{2}{*}{10}             & \texttt{Utils.java  >  isProfileOf}                            \\
                                & \texttt{SecondaryUserController.java  >  setSize}              \\ \hline
\multirow{2}{*}{12}             & \texttt{Utils.java  >  getManagedProfile}                      \\
                                & \texttt{AccountPreferenceBase.java  >  onAccountsUpdate}       \\ \hline
\multirow{4}{*}{15}             & \texttt{Utils.java  >  getBatteryPercentage}                   \\
                                & \texttt{PowerUsageSummary.java  >  onCreate}                   \\
                                & \texttt{PowerUsageSummary.java  >  onPause}                    \\
                                & \texttt{PowerUsageSummary.java  >  onResume}                   \\ \hline
\end{tabular}
\end{table}

\section{Early Results}
To demonstrate the promise of our approach, we present the results of our recommendation system on a large-active file from the Android operating system. The \texttt{Utils.java}\footnote{See \texttt{src/com/android/settings/Utils.java}.} source file of the Settings\footnote{See \url{https://android.googlesource.com/platform/packages/apps/Settings}.} application is the third most depended-on file in the system with 243 clients and the sixth most changed file with 271 commits over a two-year period.\footnote{Ending at \texttt{tags/android-12.0.0\_r3}.} Our approach successfully recommends cohesive refactoring operations that would reduce the influence of the large-active file and potentially enable further developer parallelism.

\subsection{Example of Interface Splitting}
We implemented the interface splitting algorithm as described in the previous section. We found the most success with the following configuration: (a) the clustering algorithm is the normalized cut spectral algorithm of  Shi and  Malik~\cite{shi:normalized-cuts-and-image-segmentation}; (b) $\theta$ is the number of clusters; (c) $g_q$ is the index of $q$th largest spectral gap~\cite{vonLuxburg2007}; (d) $q_{\text{max}}=3$; (e) the client set~$d(t)$ of any target function~$t$ includes both client files that directly call~$t$ and those that call~$t$ indirectly via a path of intermediate functions.\footnote{All intermediate functions must also be inside the target file.}

In Table~\ref{table:is-recs}, we present some of the more interesting recommendations for splitting \texttt{Utils.java}. Each row in this table is a recommendation to pull these functions out into their own file; otherwise known as an ``extract class'' refactoring~\cite{Fowler-2018-Refactoring}.

% Finally, our algorithmic approach frees the developer of combing through hundreds or thousands of related files and functions to devise an optimal refactoring. Instead, the rote work is done for them by a machine and they can simply browse through a list of suggestions.

The promise of this approach is that it frees the developer from having to comb through hundreds of related files and functions to determine an ideal decomposition of a large-active file. Notice how each recommendation corresponds to a clear responsibility of the target file. This is evident in the highly related function names despite lexical information not being used as input. In a large-active file, which functions correspond to which responsibilities are far from obvious. This approach saves the developer's effort by providing principled guidance on how to decompose the target file.

For instance, consider Table~\ref{table:is-clientset} which contains the clients of Recommendation 3 from the first table. Clearly, the algorithm has identified the biometrics responsibility of~\texttt{Utils.java} and has marked it for extraction because it is change-prone. While this pattern is self-evident when isolated and presented as it is here, noticing this unassisted is a far more challenging task. This approach gives the developer this ability on-demand.

A unique advantage of our approach over existing refactoring systems (discussed in Section~\ref{related-work}) is that, because we sort by change frequency, we are ensuring that the most highly ranked recommendations are those portions of the file that are  incurring the highest maintenance penalties. We avoid recommending refactorings that look attractive purely when looking at the program structure. This avoids introducing needless code churn that would result from refactoring inactive code.

Another novelty of our approach is that we offer a complete ``extract class'' refactoring as a single recommendation instead of dispersing the recommendations across a series of  ``move method'' operations. The advantage of this is that a developer can immediately see the end result and make a decision without having move forward and backward through a sequence of steps.
% Our approach offers more transparency and gives more control to the developer.

\subsection{Example of Interface Redrawing}
We implemented the interface redrawing algorithm with a similar configuration to our interface splitting implementation. The only difference is that we use the co-clustering algorithm of  Dhillon~\cite{dhillon2001co} which is also based on the spectral algorithm of Shi and Malik~\cite{shi:normalized-cuts-and-image-segmentation}.

In Table~\ref{table:ir-recs}, we present some of the more interesting recommendations for redrawing \texttt{Utils.java}. Each row in this table is a recommendation to either move these functions into their own file or to move these functions into a relevant client file.  We do not yet make the distinction between these cases and instead leave it to the developer to decide. The difference from interface splitting is that we offer functions from both sides of the interface and it is powered entirely by change history.
Each of these recommendations is a set of functions that are frequently changing together across the boundary of an interface. By recommending them, we are encouraging the developer to look closely for possible implicit coupling and consider moving them onto the same side of the interface.

No existing approach can locate the specific regions of an interface that are unstable. This provides the antidote to the unstable interface anti-pattern~\cite{Mo-2021-ArchitectureAnti-Pa} by generating concrete refactoring advice.

% This has the advantage of giving the developer more transparency and ultimately more control over the process.

% recommend extracting complete sets of functions instead of a sequence of ``move method'' recommendations. This has the advantage of allowing the developer 

% This is one core advantage our approach has over refactoring systems.

% the algorithm is able to highlight subsets of the target file that each have a clear responsibility that is evident from the function names, despite lexical information not being used as input.

% 

\section{Discussion}
This refactoring recommendation approach has the potential to increase the parallelism of developer resources. Large and overactive files are common locations to discover inappropriate coupling between components. This coupling requires careful coordination between developers to work around. The techniques we have outlined in this paper are a first step towards removing these obstacles so developers can work more independently on their desired tasks.

% This refactoring recommendation approach has the potential to dramatically reduce coupling in and around large-active files. By breaking off the responsibilities of large-active files, developers create a more maintainable and more semantic system. This has the benefit of 
% increase the parallelism of developer resources.  This coupling requires careful coordination between developers to work around. The techniques we have outlined in this paper are a first step towards removing these obstacles so developers can work more independently on their desired tasks.

A major novelty of this refactoring recommendation approach compared to existing work is its use of revision history. Each recommendation has been empirically shown to require more maintenance than the rest of the system. This is advantageous firstly because developers avoid introducing unnecessary code churn to repair code that is not causing problems. Some files might, for example, have alarming metric values but if they seldom change and incur few bugs they are not in fact problematic for the project. Secondly, the use of revision history to identify recommendations reveals instances of implicit coupling that would be missed by systems that do not consider revision history. Our approach recommends making these implicit dependencies explicit.

Another advantage of this approach is its focus on resolving the specific problem of large-active files rather than the generalist approach taken by other methods which aim to improve global quality metrics. Large and overactive files have repeatedly been shown to be a source of technical debt and a hurdle to developer parallelism. Our approach directly resolves this issue by offering complete recommendations that would narrow the responsibilities of the target file and improve the stability of its interface.

Because our approach is not global, it is suitable for both root canal and floss-style refactorings~\cite{Murphy-Hill-2008-RefactoringToolsF}. Developers may set aside some time to completely remove a particular large-active file by repeatedly querying for recommendations until there is nothing left of the target file. Or they may simply act upon a single recommendation when they are working on a file. For instance, they may choose to extract the biometric functions out of \texttt{Utils.java} when working on biometric code. Our approach is flexible enough to fit into any developer workflow.

% Importantly, this is not an automated solution. The developer is still responsible for choosing the best option (perhaps none of them.) An algorithmic approach cannot be privy to the many business decisions and domain concerns that are not represented in the data. Instead, the developer may choose to use these suggestions as a starting point for their own solution. This exploratory nature is a key advantage of our approach.

% on large-\emph{active} files. While not their focus, existing approaches may be able to decompose a large file. However, no existing

However, our approach suffers from a limitation that prevents it from practical use in its current state: it does not filter out illegal recommendations. First, it is possible that an ``extract class'' recommendation will create a cyclic dependency between the target file and the newly created file if taken by the developer. Secondly, our approach does not consider constraints introduced by the type system. Our algorithms will recommend refactorings that would break an interface implementation or a class inheritance if taken. We are currently working on solutions to these limitations.

\section{Related Work}\label{related-work}
Automated and semi-automated refactoring has received considerable attention in the field of search-based software engineering (SBSE.) An influential approach taken by  Harman and  Tratt~\cite{Harman-2007-Paretooptimalsearc} is to optimize for two global quality metrics: coupling and cohesion.  Ouni et al.~\cite{Ouni-2013-Maintainabilitydefe} and  Mkaouer et al.~\cite{Mkaouer-2015-Many-ObjectiveSoftw} continued this line of research by introducing new metrics that address the shortcomings of Harman and Tratt. Later, semi-automated approaches were introduced by Mkaouer et al.~\cite{Mkaouer-2014-Recommendationsyste},  Alizadeh et al.~\cite{Alizadeh-2019-RefBotIntelligent}, and  Lin et al.~\cite{Lin-2016-Interactiveandguid}. These improved on prior work by allowing the developer to give feedback to the system while refactoring. To our knowledge, our approach is the first to directly target large-active files.

% design anti-patterns such as cliques or unstable interfaces.

\section{Future Plans}
% We have two major plans moving forward: (a) improving our technique and (b) evaluating our approach with industrial collaborators.
As discussed, while our novel approach has the substantial promise of addressing the limits of prior refactoring systems, it is not yet suitable for practical use. Some of the refactorings recommended are illegal. Broadly, this can be remedied by either only searching through legal refactorings, or by tweaking illegal refactorings until they are legal. Further research is required to determine the most suitable method.

Once our approach is ready for practical use, we plan to evaluate our techniques with our industrial partners. Our claim of promise rests on three points: (a) our approach clarifies the responsibilities of the members of large-active files, (b) our approach is flexible enough to fit into any workflow, and (c) our approach reduces future maintenance costs by limiting or even removing the influence of the most debt-laden files. We plan to evaluate the first two points through interviews and extended collaboration with teams. The third point can be evaluated quantitatively through controlled experiments.

In the near future, we are interested in extending this research by using  lexical information in the program to recommend names for the newly extracted files. This would further explain the responsibilities of the recommended files and improve the usability and adoption of our approach.

% - flexibility
% - reduces future maintenance costs
% - clarifies the responsibilities of files surrounding large-active files

% We believe a core advantage of our approach over existing work is that it is flexible enough to fit any kind of workflow.

% To address, we plan to constrain our clustering algorithm to only search
% \begin{enumerate}
%     \item A full empirical evaluation, quantitative and qualitative
%     \item Address the illegal refactoring limitation
%     \item Interface redrawing should be more specific
%     \item Recommend names for the newly extracted files
% \end{enumerate}

\section{Conclusion}
Complex, debt-laden code does not appear overnight. It is created over  months and years of gradual erosion of the design. Large-active files are clear examples of this erosion. We have introduced a novel method of decomposing large-active files by using a project's revision history and the shared dependents of functions within the file to form recommendations. Each recommendation highlights a single responsibility. By moving responsibilities of the large-active file out and into their own files, we reduce the overall coupling of the system. When coupling is reduced, developers are more able to work more effectively in parallel and the overall complexity of the system is reduced, which should lead to fewer bugs and simpler changes.

Our approach is the first of its kind to directly target large-active files and decompose them using their influences. The key advantage of our approach is its use of revision history to ensure recommendations isolate meaningful technical debt. Although our approach is not yet fully-featured, we plan to address its limitations and provide a full evaluation of these techniques soon.

\AtNextBibliography{\small}
\printbibliography
% \end{thebibliography}

\end{document}